%% file: paper-ieee.tex
\documentclass[conference]{IEEEtran}
\IEEEoverridecommandlockouts
\pdfoutput=1
\usepackage{cite}
\usepackage{amsmath,amssymb,amsfonts}
\usepackage{algorithmic}
\usepackage{graphicx}
\usepackage{textcomp}

\usepackage[urlcolor= blue]{hyperref}
\usepackage{xcolor}
\def\BibTeX{{\rm B\kern-.05em{\sc i\kern-.025em b}\kern-.08em
    T\kern-.1667em\lower.7ex\hbox{E}\kern-.125emX}}
    
\usepackage{multirow}
\usepackage{tabularx, booktabs}
\usepackage{hhline}
\usepackage{multicol}
\usepackage{enumitem}
\usepackage{lipsum}
\usepackage{amsmath}
\usepackage{flushend}

\newcommand{\cameraready}[1]  {{#1}}

 \newcolumntype{L}[1]{>{\raggedright\let\newline\\\arraybackslash\hspace{0pt}}m{#1}}
\newcolumntype{C}[1]{>{\centering\let\newline\\\arraybackslash\hspace{0pt}}m{#1}}

\AtBeginDocument{%
  \providecommand\BibTeX{{%
    \normalfont B\kern-0.5em{\scshape i\kern-0.25em b}\kern-0.8em\TeX}}}
    
    \newenvironment{boxedtext}
    {
    
    \begin{center}

    \begin{tabular}{|p{0.96\linewidth}|}
    \hline
    }
    { 
    \\ \hline
    \end{tabular} 
    
    \end{center}
    \vspace{5pt}
       }

\begin{document}


\title{ToxiSpanSE: An Explainable Toxicity Detection in Code Review Comments}


\author{\IEEEauthorblockN{Jaydeb Sarker$^\clubsuit$, Sayma Sultana$^\clubsuit$, Steven R. Wilson$^\diamondsuit$, Amiangshu Bosu$^\clubsuit$}
\IEEEauthorblockA{\textit{$^\clubsuit$Wayne State University, Detroit, Michigan, USA} \\
\textit{$^\diamondsuit$Oakland University, Rochester, Michigan, USA}\\
\emph{jaydebsarker@wayne.edu}, \emph{sayma@wayne.edu}, \emph{stevenwilson@oakland.edu}, \emph{amiangshu.bosu@wayne.edu}} 
}


\maketitle
\IEEEpubidadjcol

\begin{abstract}
\input{Sections/abstract}
\end{abstract}


\input{Sections/introduction}
\input{Sections/background}

\input{Sections/research-method}

\input{Sections/evaluation}

\input{Sections/discussion}

\input{Sections/threats}
\input{Sections/conclusion}

\bibliographystyle{IEEETran}
\bibliography{toxicity-references}

\end{document}

%% file: Sections/abstract.tex
\cameraready{Background:} The existence of toxic conversations in open-source platforms can degrade relationships among software developers and may negatively impact software product quality. To help mitigate this, some initial work has been done to detect toxic comments in the Software Engineering (SE) domain.

\cameraready{Aims:} Since automatically classifying an entire text as toxic or non-toxic does not help human moderators to understand the specific reason(s) for toxicity, we worked to develop an explainable toxicity detector for the SE domain.

\cameraready{Method:} Our explainable toxicity detector can detect specific spans of toxic content from SE texts, which can help human moderators by automatically highlighting those spans. This toxic span detection model, \textit{ToxiSpanSE}, is trained with the 19,651 code review (CR) comments with labeled toxic spans. Our annotators labeled the toxic spans within 3,757 toxic CR samples. 
We explored several types of models, including one lexicon-based approach and five different transformer-based encoders. 

\cameraready{Results:} After an extensive evaluation of all models, we found that our fine-tuned RoBERTa model achieved the best score with 0.88 $F1$, 0.87 precision, and 0.93 recall for toxic class tokens, providing an explainable toxicity classifier for the SE domain. 

\cameraready{Conclusion: Since \textit{ToxiSpanSE} is the first tool to detect toxic spans in the SE domain, this tool will pave a path to combat toxicity in the SE community.}

%% file: Sections/introduction.tex
\section{Introduction}
\label{sec:intro}
Toxicity, which is a large umbrella term comprising various antisocial behaviors such as offensive language, cyberbullying, hate speech, and sexually explicit content~\cite{barbarestani2022annotating}, is pervasive among various online platforms~\cite{miller2022did,anderson2018toxic}.  
As most of the Free and Open Source Software (FOSS) communities operate online, they are not immune from such toxic interactions~\cite{ramanstress,miller2022did,sarker2023automated}.
As software development requires close collaboration and rapport among participants, toxicity can have severe repercussions for a FOSS community, which include decreased productivity, wastage of valuable time~\cite{ramanstress}, negative feelings among the participants~\cite{egelman2020predicting}, barriers to newcomers' onboarding~\cite{steinmacher2014support,jensen2011joining}, hostile environments towards minorities~\cite{gunawardena2022destructive}. As proactive identification and mitigation of toxic interactions among FOSS developers are crucial, automated approaches can help FOSS moderators.

\cameraready{Prior studies~\cite{sarker2023automated,sarker2020benchmark} found that off-the-shelf toxicity detectors do not perform well in the SE texts because some words (`die', `kill', `dead') in the SE context have a different meaning.} Due to the unreliability of off-the-shelf natural language processing (NLP) tools on Software Engineering (SE) datasets~\cite{sarker2020benchmark,jongeling2017negative}, recent works have proposed customized toxicity detectors trained on SE communications~\cite{ramanstress,sarker2023automated}. While these tools boost reliable performances on SE datasets, we have identified a shortcoming of these two solutions. First, existing tools classify an entire paragraph on a binary scale, including hundreds of sentences. Even if only one of those sentences is toxic, it classifies the whole paragraph as toxic. 
A binary, paragraph-level classification of toxic texts may help the FOSS community to decide to remove a particular paragraph or establish a code of conduct for toxic comments. However, it becomes time-consuming for a moderator to identify the offending excerpt(s) from a large paragraph. \cameraready{Second}, due to the lack of cultural differences, a moderator may fail to identify the offending sentences from a paragraph classified as toxic by these tools. Being motivated by recent advances in explainable machine learning (ML) models, this study aims to create a new  SE domain-specific toxicity detector that overcomes this particular shortcoming. We aim to  \textit{develop an explainable toxicity detector for the Software Engineering domain, which can precisely identify toxic excerpts from a text to assist FOSS moderators.}
\cameraready{`Explainable' in the context of this study indicates the ability of the classifier to pinpoint the words/phrases responsible for a text's toxic classification~\cite{ribeiro2016should}.}

 Our solution aims to pave a path for automated text moderation to foster healthy and inclusive communication by reducing manual efforts to locate the toxic contents in FOSS developers' communication and helping project maintainers quickly identify the negative parts of the comment to decide whether the text should be approved or rejected. Moreover,
 \cameraready{this technique} will also enable finer-grained toxicity analyses from the patterns of toxic excerpts to determine possible remedies. Finally, our work can be a building block to develop solutions to proactively prevent toxic communications, similar to grammatical mistakes/typos detection tools.

A \textit{toxic span} is defined as the fragment of a sentence or text that potentially causes the meaning of the text to be toxic~\cite{pavlopoulos2021semeval}. A toxic span may contain a single word or a sequence of words. For example, ``\textcolor{red}{Yuck}, this code is a \textcolor{red}{crap}'' where the toxic spans are highlighted with \textcolor{red}{red} color. 
The SemEval-2021's Task 5 organizers provided 10K toxic posts from the Civil Comments dataset~\cite{borkan2019nuanced} with labeled span character offsets. An ensemble solution using BERT~\cite{devlin2018bert} achieved the best performance among the teams participating in this challenge.
As prior research shows the necessity of SE domain-specific customization for NLP tools~\cite{jongeling2017negative,sarker2020benchmark}, these toxic span detectors may not perform well on SE texts. Hence, we aim to build a customized solution.

On this goal, we select SE domain-specific toxicity dataset from Sarker \textit{et} al.~\cite{sarker2023automated}, which consists of a total of 19,651 \cameraready{Code Review (CR)} comments with 3,757 ($\sim19\%$) toxic samples. We manually label this dataset using two independent raters to develop ground-truth annotations for the toxic spans within the toxic samples. We \cameraready{measured} inter-annotator agreement using Krippendorff’s $\alpha$~\cite{krippendorff2004reliability}, which was 0.81 (almost perfect agreement).
We first developed a lexicon-based classifier using this dataset to establish a baseline model. We trained and evaluated five sequence-to-sequence transformer models. During our 10-fold cross-validation-based evaluations, we found a model based on a fine-tuned RoBERTa~\cite{liu2019roberta} achieving the best an $F1-$score of 0.88. Primary contributions of this work include:
\begin{itemize}
    \item \textit{ToxiSpanSE}: The first explainable toxicity detector for the SE domain. 

    \item An expert-annotated, span-level toxicity labels for 3,757 toxic code review comments. 
    
    \item An overview of metrics to develop explainable NLP tools for the SE domain.
    
    \item An empirical evaluation of five transformer-based models with 19,651 code review texts. 
    \item We make our model and dataset available for further analysis and use in the software engineering community. Available at: 
    \url{https://github.com/WSU-SEAL/ToxiSpanSE}
\end{itemize}

\textbf{Paper organization:} The remainder of this paper is organized as follows. Section~\ref{background} provides the related works on toxicity and toxic span detection. We discuss the research methodology in Section\ref{method}. Section~\ref{sec:evaluation} presents the results. Section~\ref{discussion} discusses the lesson learned. Section~\ref{threats} addresses the threats to validity of this work. Finally, Section~\ref{conclusion} concludes the paper.

%% file: Sections/background.tex
\section{Background}
\label{background}

\subsection{Toxicity Phenomena}

The term `toxicity’ represents the negative or antisocial interactions in online conversations~\cite{aroyo2019crowdsourcing}. A report from ~\cite{aroyo2019crowdsourcing} showed that 47\% Americans experienced harassment and abuse during online communication, and toxicity deters users from online engagement.
Toxicity is a subjective phenomenon often subject to the opinions of beholders~\cite{lenhart2016online}. A broader view of toxicity is that of an umbrella of various antisocial behaviors such as hate speech, cyberbullying, trolling, and flaming~\cite{miller2022did}. The Conversational AI team from Google defined toxicity as ``comments that are rude, disrespectful or otherwise likely to make someone leave a discussion''~\cite{perspective-api}.

Toxicity in the SE domain is not uncommon. Recent studies from the SE domain analyzed toxicity and other antisocial behaviors. Sarker \textit{et} al. defined a code review comment as toxic if it includes any antisocial behaviors such as offensive name-calling, insults, threats, personal attacks, flirtations, sexual reference, and profanities~\cite{sarker2020benchmark}. Miller \emph{et} al. adjusted the meaning of toxicity from Conversational AI during analyzing GitHub issue discussions and mentioned that the toxicity umbrella covers trolling, flaming, hate speech, harassment, arrogance, entitlement, and cyberbullying~\cite{miller2022did}. Ferreira \emph{et} al. defined the unnecessary disrespectful term toward discussion as incivility during analyzing the Linux Kernel Mailing list~\cite{ferreira2021shut}. A similar term of toxicity is `destructive criticism’ during code review~\cite{gunawardena2022destructive} that represents the negative feedback, including threats, poor task performance, or flaws of the individuals. In 2020, Egelman \emph{et} al. defined `pushback’ in code review as a reviewer blocking a change request due to unnecessary interpersonal conflict~\cite{egelman2020predicting}. In this work, we adapted the definition of toxicity from the study of Sarker \emph{et} al.~\cite{sarker2020benchmark} because we used their comment-level toxicity labels, which were annotated using this definition, and they provided a rule book for marking a text as toxic or non-toxic which assisted us with our annotation process.

\subsection{Toxicity in SE}
FOSS communities have reported toxic contents in developers' communications in blog posts~\cite{toxic-blog-linux1, toxic-blog-linux2}, podcasts~\cite{toxic-podcast}, and talks~\cite{toxic-talk}. Large open-source foundations face an increase in toxicity in their groups. For example, the Linux Community experiences toxicity~\cite{toxic-blog-linux2}, and the founder apologized for using toxicity in Linux Kernel Mailing lists~\cite{toxic-blog-linux3}. By conducting a survey, the Perl Foundation found that several members stepped down due to receiving abusive messages~\cite{perl-toxicity}.

Researchers from the SE community also conducted empirical studies to understand toxicity in FOSS domain~\cite{carillo2016towards, squire2015floss}. Another line of research focused on analyzing the toxic behavior of open source developers' interactions such as GitHub issue discussions~\cite{ramanstress, miller2022did}, code review comments~\cite{sarker2020benchmark,sarker2023automated, qiu2022detecting, gunawardena2022destructive}, and Gitter messages~\cite{sarker2020benchmark}. 
Toxic communication impacts developers' mental health like a `poison'~\cite{carillo2016towards} and can cause burnout~\cite{ramanstress}. To combat toxicity in open source, Raman \emph{et} al. developed a toxicity classifier trained with a small-scale GitHub issue discussion dataset~\cite{ramanstress}. However, several studies showed that their tool~\cite{ramanstress} performed poorly on the large-scale SE texts~\cite{sarker2020benchmark, qiu2022detecting, miller2022did, sarker-ase22}. The number of annotated toxic texts in open-source communication is relatively small, which makes it challenging to build a reliable toxicity classifier. However, the existence of toxicity has severe repercussions among the open-source software developers such as newcomers onboarding~\cite{ramanstress}, disproportionate impacts on underrepresented groups~\cite{sarker2022identification, gunawardena2022destructive, ramanstress}, and  pushbacks~\cite{egelman2020predicting}. 
To detect the toxicity from code review comments, Sarker \emph{et} al. developed a machine learning-based toxicity classifier (referred to as `ToxiCR') trained with 19,651 labeled code review comments from Gerrit projects~\cite{sarker2023automated}. In 2022, 
Qiu \emph{et} al. developed a classifier for identifying interpersonal conflicts during code reviews~\cite{qiu2022detecting}. Cheriyan \emph{et} al. developed a classifier to detect swearing and profanity~\cite{cheriyan2021towards} for four different SE platforms. 
Our work differs from the prior works by focusing on finer-grained identification of which phrases make a text toxic in the SE context.

Recent studies also focused on understanding antisocial behaviors in open-source communities. Miller \emph{et} al. conducted a study for a better understanding the toxicity with a qualitative analysis of 100 toxic issue comments on GitHub~\cite{miller2022did}. They manually investigate the nature, context, participants, and after impacts of toxicity on GitHub. After analyzing the 1,545 emails from Linux Kernel Mailing lists, Ferreira \emph{et} al. found that the common forms of incivility are frustration, name-calling, and \cameraready{impatience}~\cite{ferreira2021shut}.  Gunawardena \emph{et} al. defined `destructive criticism' as another antisocial behavior, which  includes negative feedback which is nonspecific and is delivered in a harsh or sarcastic tone. Their survey of 93 developers suggests destructive criticism as a barrier to promoting diversity and inclusion~\cite{gunawardena2022destructive}. 

\subsection{Toxic Span Detection}

Although the detection of toxicity~\cite{pavlopoulos2017deeper, bhat2021say, georgakopoulos2018convolutional}, hate speech~\cite{burnap2014hate, burnap2015cyber, gitari2015lexicon}, and offensive language~\cite{chen2012detecting, isaksen2020using} are common in online platforms, the idea of span detection of toxicity has only more recently gained attention with  the SemEval-2021 toxic span detection task~\cite{semEval2021toxic}. Toxic spans represent a part of the text that is responsible for the toxicity of the posts~\cite{pavlopoulos2021semeval}. This direction is inspired by prior NLP studies on aspect-based sentiment analysis~\cite{xu2020aspect, qiang2020toward}, which aims to detect the sentiment of a text and find the specific region of a text that expresses the sentiment using attention-based deep neural network models~\cite{xu2020aspect}. While earlier studies focused on the attention mechanism championed by Vaswani \textit{et} al.~\cite{vaswani2017attention}, Sen \emph{et} al. found that machine attention does not reliably overlap with human attention maps~\cite{sen2020human}. 
To improve explainability using attention-based  mechanisms, recent works have proposed transformer-based sequence-to-sequence models~\cite{chefer2021transformer, qiang2022attcat}. 

In the SemEval-2021 task, \cameraready{Pavlopoulos \textit{et al.}} provided a labeled dataset~\cite{semEval2021toxic} of toxic spans with 10,000 samples~\cite{pavlopoulos2021semeval} curated from the Civil Comments dataset. The raters marked the span that corresponds to the toxicity of a text. The task is a binary classification because it contains toxic and non-toxic tokens. Moreover, they fixed the ground truth of the dataset if the majority of the raters labeled the span as toxic. Ninety-one teams made submissions with different methods in the SemEval-2021 competition to detect toxic spans~\cite{semEval2021toxic}. One of the teams proposed a BERT-based ensemble method toxic span detection approach where they achieved 70.83\% $F1$ score and secured first place in SemEval-2021 task~\cite{zhu2021hitsz}. 
A RoBERTa-based method performed only slightly worse, with a 70.77\% $F1$ score, and other approaches based on fine-tuning of pre-trained transformer models (\cite{suman2021astartwice, chhablani2021nlrg}) also performed well for that toxic span detection tasks. Since several studies worked on toxic span detection for online civil comments, Pavlopoulos \emph{et} al. annotated a new dataset for toxic to civil transfer~\cite{pavlopoulos2022detection}. Although several studies have proposed toxic span detectors for online comments, no such tool exists for the SE domain. Since NLP tools may not work reliably on a cross-domain dataset~\cite{jongeling2017negative}, the development and evaluation of a SE domain-specific toxic span detector are essential. Such a tool will not only enable a finer-grained analysis of toxicity but also enable proactive notification to authors.

%% file: Sections/research-method.tex
\section{Research Method} \label{method}

After selecting a dataset from a prior work~\cite{sarker2023automated}, two of the authors independently annotated toxic spans in each text. Using this annotated dataset, we train and evaluate sequence-to-sequence transformer models that output the probability of each word belonging to a toxic span in the current text context. Finally, we use postprocessing steps to identify toxic spans from the output probabilities based on empirically determined thresholds. The following subsections detail our research methodology.

\subsection{Dataset}

\subsubsection{Dataset Selection}

The number of datasets for toxicity detection in Software Engineering communication is small~\cite{ramanstress, sarker2020benchmark}. We explored previous studies on toxicity and antisocial behaviors in open-source interactions and found four studies that provided manually labeled datasets for toxicity~\cite{ramanstress, sarker2020benchmark, sarker2023automated} and incivility~\cite{ferreira2021shut} detection. Raman \emph{et} al. labeled only 611 texts from GitHub issue discussions as toxic or non-toxic~\cite{ramanstress}. In 2020, Sarker \emph{et} al. provided a dataset of 6,533 CR comments and 4,140 Gitter messages labeled as toxic or non-toxic~\cite{sarker2020benchmark}. They also provided a rubric to identify a text as toxic or non-toxic. In a subsequent study of building a toxicity detection tool, they annotated 19,651 CR comments with binary, comment-level toxicity scores~\cite{sarker2023automated}. Moreover, Ferreira \emph{et} al. provided an annotated 1,545 emails from the Linux Kernel Mailing List where they labeled each message as civil or uncivil~\cite{ferreira2021shut}. Given that Sarker \emph{et} al.’s dataset~\cite{sarker2023automated} is the largest one in the Software Engineering domain for toxicity detection,  we select their dataset for our study. Moreover, their detailed rubric also guides our annotators on how to label toxic spans.  

\subsubsection{Dataset Annotation}

We got each of our toxic samples manually annotated by two independent annotators. To diversify the annotators, we chose one woman and one man for the annotation task. As manual labeling toxic text is a subjective task, we sought to reduce subjectivity bias during manual annotation by asking our annotators to carefully read and follow the rubric for toxicity developed by~\cite{sarker2023automated}. Although Sarker \emph{et}  al.'s dataset~\cite{sarker2023automated} includes  19,651 CR comments, only 3,757 are labeled as toxic. Therefore, our annotators only labeled the 3,757 toxic ones, assuming that the non-toxic samples do not include any toxic spans (empty span offsets).

\begin{figure}
	\centering  \includegraphics[width=9 cm]{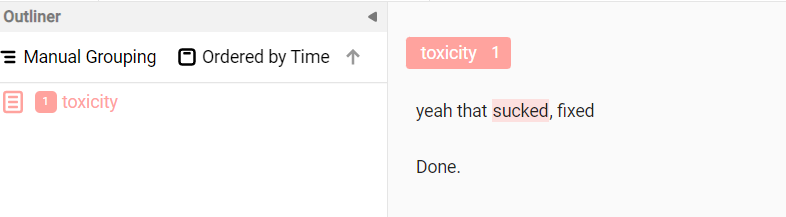}
\caption{Manual Labeling using Label Studio, toxic span is \textcolor{red}{highlighted}}
\label{fig:labeling}	
	
\end{figure}

For annotation, we use the Label Studio platform~\cite{Label}. Figure~\ref{fig:labeling} shows an example of our annotation interface. We exported the labeled data from Label Studio, which returns the code review text and corresponding character span offsets of the toxicity annotations for each sample. Table~\ref{tab:span_offset} shows two example annotations. The first example shows a toxic sample where the word `sucked' makes text toxic, and this span occurs in character offsets 10-15. The third example is non-toxic and therefore has no span selected, resulting in an empty list ([]) of character offsets.

 \begin{table}
    \caption{{Raw dataset with character spans.} {\textcolor{red}{Red}{ marked represents selected toxic words}}}
    \label{tab:span_offset}
    \centering
    \input{Tables/span_offset}
\end{table}

\subsubsection{Inter-annotator Agreement}

\begin{table*}
    \caption{{Example of Inter-Rater Agreement and Conflict Resolution}}
    \label{tab:labeling}
    \centering

\input{Tables/labeling}
    \vspace{-12pt}
\end{table*}

We wrote a Python script to compare the spans produced by the two annotators. Unsurprisingly, we have found conflicts between the labeling samples. Previous studies have suggested several \textit{chance-corrected} agreement measures to compute the inter-annotator agreement (IAA)~\cite{braylan2022measuring}. \textit{Chance-corrected} measures such as Cohen's $\kappa$ ~\cite{cohen1960coefficient}, Fleiss' $\kappa$~\cite{fleiss1971measuring}, and Scott's $\pi$~\cite{scott1955reliability} distinguish the observed disagreements (${D}_{o}$) from expected disagreements ($D_{e}$).  Therefore, these IAA measures are unsuitable for sequential tagging with potential partial overlaps~\cite{mathew2021hatexplain}. 

Hence, similar to prior studies developing sequence tagging datasets~\cite{mathew2021hatexplain, ousidhoum2019multilingual, del2017hate, braylan2022measuring}, we chose Krippendorff’s $\alpha$~\cite{krippendorff2004reliability} as the IAA measure. 
Krippendorff’s $\alpha$ is more robust as it can handle multiple annotators and missing values and considers partial agreement/disagreements among the labelers. Krippendorff’s $\alpha$ allows the distance-based formulation and it is designed for context-specific tasks. The formula of Krippendorff's is: $\alpha=1-\frac{\hat{D}_{o}}{\hat{D}_{e}}$, for a given distance function of $D(a,b)$ where $\hat{D}_{o}$ represents the observed average distance and $\hat{D}_{e}$ is expected average distance~\cite{braylan2022measuring}. Since Krippendorff’s $\alpha$ calculates several distance functions such as nominal, interval, and ordinal~\cite{krippendorff2004reliability}, we chose the nominal distance function for our measurement. To calculate Krippendorff’s $\alpha$ score, we wrote our script using the existing implementation~\cite{krippendorff-alpha}.

Our labeled dataset has 3757 toxic code review samples labeled by two raters for toxic spans. For calculating Krippendorff’s $\alpha$ with nominal distance, created two arrays of labels. We split each sample ($s$) to a set of tokens $s= t_{0},t_{1},.....,t_{j}$ where $t_{j}$ is a token inside the sample $s$. 
As our primary dataset contains the character level span offsets, we preprocessed it for token-level offsets. Table~\ref{tab:labeling} shows an example of defining the token array for a sample. There is a total of 15 tokens after excluding the comma (,) from the input text. Therefore, we generate an array of 15 elements (same as the length of tokens) in which each position corresponds to a token from the CR text. We have a same-length array for Rater1 and Rater2 where we set 1 if the token is inside the span selection, otherwise 0.
Following this process, we generated 3,757 arrays for all toxic samples of Rater1 and Rater2. For computing agreement, we merge all the token-level annotations for each rater into a single array where each array contains a total of 84,951 ratings. We calculated Krippendorff’s $\alpha$ using the nominal distance between these two arrays and found the $\alpha$ value as 0.81 (almost perfect agreement). This agreement score is significantly higher than a prior work~\cite{mathew2021hatexplain} where the agreement score $\alpha$ is $0.46$.

\subsubsection{Conflict Resolution and Ground Truth}

We found that two labelers have at least partial disagreement in 928 samples. Two of our raters (Rater1 and Rater2) discussed resolving the conflicts and assigned the final labels. Table~\ref{tab:labeling} shows an example conflict with token arrays and corresponding character spans to illustrate our resolution process. 
At the end of this step, our final dataset includes CR comments and the corresponding character spans.

\subsection{Tool Design}

\begin{figure}
	\centering  \includegraphics[width=0.92\linewidth]{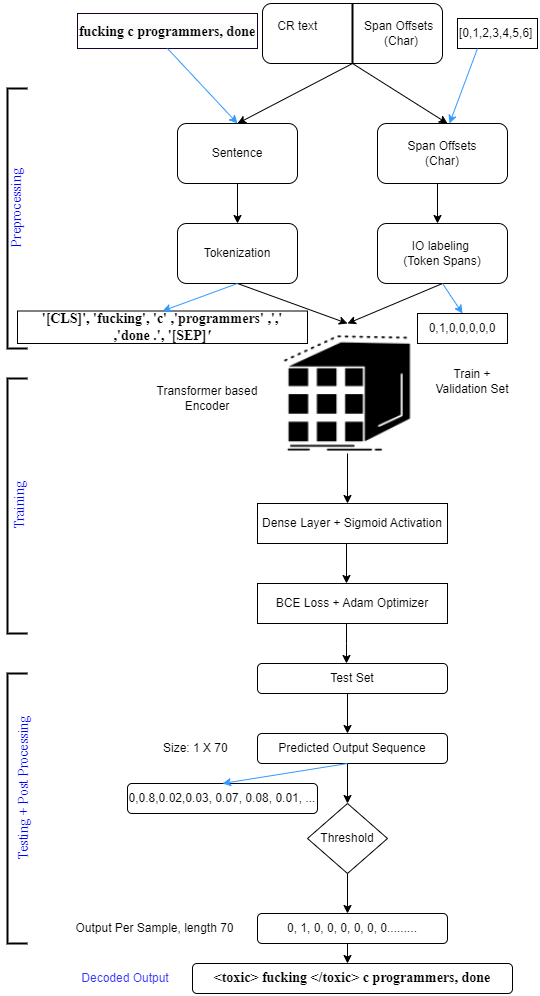}
\caption{{Model Architecture of ToxiSpanSE. Optimal threshold for each model was empirically selected (detailed in~~\ref{sec:threshold-selection})}, {\textcolor{blue}{Blue arrow $\rightarrow$}  shows an example}}
\label{fig:model_arch}	
	
\end{figure}

\cameraready{We compared two different approaches to design \textit{ToxiSpanSE}}. 
First, we used a lexicon-based naive approach, where words belonging to a predefined list are marked as toxic spans. Second, we used a supervised learning-based approach with five different transformer-based encoders. Figure~\ref{fig:model_arch} depicts our model architecture for the transformer-based models with an example prediction. 

\emph{ToxiSpanSE} takes input texts and associated labeled spans as input. After preprocessing, inputs are passed to the transformer models. The output of those models are arrays of floating point numbers ranging from 0 to 1, which indicate the probability of each token belonging to a toxic span. The following subsections detail 
lexicon-based and transformer-based approaches.

\subsubsection{Preprocessing}
The model takes the CR text and the target spans (\cameraready{labeled spans}) with character offsets as input. Further, we split each text into sentences using en\_core\_web\_sm from the spacy library~\cite{spacy2} and keep corresponding character span offsets for each sentence. We have 39,438 sentences after splitting 19,651 CR texts; among those, 5,465 sentences have toxic spans. Therefore,  around $13.85\%$ samples in our dataset have at least one toxic span. We chose sentence-level evaluation for two reasons: 
i) \cameraready{a sentence may itself have} toxic spans, ii) prior work also did sentence splitting for toxic span detection~\cite{suman2021astartwice}. 
Further, we apply a tokenizer to convert each sentence to corresponding tokens. In this study, we use tokenizers that are appropriate for each model. 
For the lexicon-based model, we chose \textit{NLTK word\_tokenize~\cite{loper2002nltk}} from Python. On the other hand, we use transformer-based encoder models' corresponding tokenizer from huggingface~\cite{wolf-etal-2020-transformers}.
We use \textit{AutoTokenizer} function and select: i) \textit{bert-base-uncased}, ii) \textit{roberta-base}, iii) \textit{distilbert-base\-uncased}, iv) \textit{albert-base-v2}, and v) \textit{xlnet-base-cased} tokenizers for their corresponding encoder model. Moreover, we set the maximum length to 70 during the tokenization of each sentence, as \cameraready{using a Python script} we empirically found that 98.5\%  of our sentence samples have less than 70 tokens. This pruning was essential as taking a large token length significantly increases both required memory and training time.   
Each transformer-based pre-trained tokenizer splits the sentences into sub-word token strings and adds an unknown token to its dictionary if it finds them. For each sentence, each transformer-based tokenizer generates a special token at the start of the sentence at the end of the sentence (i.e., the bert-base tokenizer puts
the [CLS] token at first and the [SEP] token at the end of each text). To pass the tokens of each sentence into the encoder model, we take three inputs for each sample from the tokenizers that are input\_ids, token\_type\_ids, and attention\_mask. We can decode the vector to the original string using  input\_ids.

\subsubsection{IO Encoding}
Prior NLP works with sequence labeling datasets followed BIO~\cite{suman2021astartwice} or IO~\cite{chen2021ynu} tagging to encode the spans. \emph{BIO} stands for Beginning, Inside, and Outside, where \emph{B}-indicates the \cameraready{beginning} token of a toxic span, \emph{I}- indicates that the token is inside the toxic span, and \emph{O} indicates a token outside the toxic span. BIO is suitable for NLP tasks to divide a span of text into multiple chunks. As we aim to identify which text spans contain something toxic, a simpler one, i.e., the IO -encoding, is sufficient for our goal. Moreover, IO simplifies our processing steps. In our IO encoding, every \emph{I} tag corresponds to a token inside a toxic span, and \emph{O} indicates outside.

To get the target span, we use their \textit{offset\_mapping} to determine whether that token is inside the selected toxic span. \textit{Offset\_mapping} provides each token's starting and ending character. Next, we generate a sequence of 0s (non-toxic token) and 1s (toxic token) for each sentence. Hence, our ground truth target is a sequence of 1's and 0’s of maximum length (70). 
We consider the first and last token value as 0 for each sample because each tokenizer of pre-trained transformers generates a special token at the start and another at the end. Finally, for each sentence, we have a vector (length = 70) containing a sequence of 0s and 1s, which is the ground truth target vector.

\subsubsection{Lexicon Based Model}

We also designed a naive model referred to as the `lexicon-based' model for detecting toxic spans from our dataset. In general, toxic spans contain many common words, including profanity, sexually explicit, and swear words. The purpose of developing this model is to evaluate whether a simple lexicon search-based approach compares against state-of-the-art transformer-based models. Our lexicon-based model matches each token in a text against a list of common toxic tokens with our ground truth tokens. We curated a list of toxic tokens  ($\Sigma TOK$ = $tok_{0}, tok_{1},....,tok_{i}$)  from two prior studies, which include  85 profane words from Sarker \emph{et} al.~\cite{sarker2023automated} and the top 100 toxic tokens from Kurita \emph{et} al.~\cite{kurita2019towards}. In total, our lexicon list contains 167 tokens since there are overlapping tokens between those two lists. 

\textit{Ground Truth For \cameraready{Lexicon Based Classification}:} To tokenize each sentence, we use \textit{NLTK word\_tokenize} from Python. Further, we use \textit{textspan} library~\cite{TextSpan} to get the exact location of selected tokens from the human labeling spans. We use a similar IO encoding approach for this model (70-length vector for each input) where  
if that is inside the labeled span, we put the token position as 1, otherwise 0.  

\textit{Lexicon Based Classifier Output:} We generate an output vector($vec$) for each input sentence with a length of 70. We set the $vec_{i}=1$ if that token of the input matches with one of the tokens from the $TOK$, and $vec_{i}=0$, otherwise.

\subsubsection{Transformer based model}
The Transformer deep learning architecture that emerged in 2017~\cite{vaswani2017attention} is based on multi-head self-attention and has shown to perform significantly better than Recurrent Neural Network (RNN)-based models for sequence-to-sequence tasks. Transformers use a self-attention mechanism for computing the internal representation of input and outputs. Moreover, the transformer-based model does not require any pre-computed context-free embedding vectors. Instead, it can generate context-based embeddings by pre-training the entire model as an encoder for sequence-to-sequence tasks.

From the preprocessing steps, we have 
inputs (input\_ids, token\_type\_ids, attention\_mask) for each sentence and we have generated \cameraready{ground truth} (target labels) using IO encoding. Inputs and targeted IO encoding are passed to the encoder layer to generate context-based embeddings. \cameraready{Since there are several Transformer based encoders available for sequence classification tasks, we consider the following transformers which performed well in a prior token-level classification task~\cite{semEval2021toxic}}. In this work, we used Transformer based encoders from the HuggingFace library~\cite{wolf-etal-2020-transformers}, selecting the following pre-trained encoders: 

\begin{itemize}
    \item {BERT}: Devlin \emph{et} al. proposed the pretraining of the Deep Bidirectional Transformers for Language Understanding (BERT) model in 2018 that was trained with masked language modeling (MLM) and next sentence prediction (NSP)~\cite{devlin2018bert}. We use the BERT-base model, which has 12 transformer layers with 768 hidden states and 12 attention heads with 110 M parameters. BERT can be fine-tuned with domain-specific datasets for sentence and sequence classification tasks. 

  \item DistilBERT: Sanh \textit{et} al. proposed a lighter and faster version of the bert-base model, using modeling distillation, referred to as ``DistilBERT’’~\cite{sanh2019distilbert}. It has around 66 M parameters (40\% 

  \item RoBERTa: An optimized version of BERT is RoBERTa, which achieved better performance than BERT-base in some NLP tasks by pretraining the model for a longer time and on more data than the original BERT~\cite{liu2019roberta}. The base model has the same architecture as the BERT-base model. 

\item ALBERT: ALBERT has a similar architecture to the BERT-base but has only 128 hidden embedding layers that reduced the total parameters to 12 M~\cite{lan2019albert}. We chose to use the ALBERT-base model for this study. 

\item XLNet: Unlike the autoencoder (AE) language models (i.e., BERT), Yang \textit{et} al. proposed XLNet, which is based on autoregressive language modeling~\cite{yang2019xlnet}. XLNet sought to overcome the limitations of the BERT model by maximizing the expected likelihood over all permutations of the factorization order. Moreover, its performance does not rely on data corruption. We use \textit{xlnet-base-cased} model from the transformer library, which has a similar size as \textit{BERT-base} model.

\end{itemize}

In this experiment, we select those pre-trained encoders from the HuggingFace library~\cite{wolf-etal-2020-transformers}.
After the embeddings with size (1 X 70), we set a Dense layer to set the required final output size. Moreover, since we are doing a binary sequence classification task, the `sigmoid’ activation function is added to this Dense layer to generate the final output's probability. 
 Therefore, our final output vector is a sequence of floating point values (from 0 to 1 due to the sigmoid function) with a length of 70.

\subsubsection{Post Processing}

After fine-tuning the model (details in next section), we predict the probability score with the test samples. The model provides a probability score from 0 to 1 for each token (70 per sample ($s$)). Using an empirically determined threshold (\cameraready{Section~\ref{sec:threshold-selection}}) parameter, we decide whether a token is in the toxic class (1) or non-toxic (0). Further, from the prediction vector, we generate a set ($Pred_{s}$) of indexes for the output tokens in the toxic class. Our ground truth has already been preprocessed as toxic and non-toxic tokens. We also generate a set of the indexes of toxic tokens from the ground truth ($G_{s}$) of the test set. Finally, we wrote a Python script to decode each token from the sample and show the output like figure~\ref{fig:model_arch}. We have input \textit{``fucking c programmers, done''}, and the model provides the output \textit{''\textless toxic\textgreater  fucking \textless/toxic\textgreater c programmers, done''}. To make the tool user-friendly, we use a tag (\textless toxic\textgreater) at the start and  (\textless /toxic\textgreater)  at the end for predicted tokens inside toxic spans.

%% file: Tables/span_offset.tex
    \begin{tabular}{|p{3cm}|p{5cm}|} \hline

     \textbf{Character Span Offsets}     & \textbf{CR Text}  \\ \hline
[10, 11, 12, 13, 14, 15] &  Yeah that \textcolor{red}{sucked}, fixed done.    \\  
\hline
[39, 40, 41, 42, 43, 44, 45, 46, 47, 48, 74, 75, 76, 77]& I think the formatting may have gotten \textcolor{red}{screwed up} (or Gerrit made it look \textcolor{red}{ugly}) \\ \hline
[ ] &  below assignments also should be removed    \\  \hline

    \end{tabular}

%% file: Tables/labeling.tex
  \begin{tabular}{|p{1.3 cm}|p{8cm}|p{4cm}|p{2.2cm}|} \hline

{\textbf{Rater}}  &{\textbf{Text}} &{\textbf{Token Array}}  & C\textbf{haracter Spans} \\ \hline

 Rater1 &  if you think it \textcolor{red}{sucks horribly}, that's fine as long as we can fix it
 
  &[0,0,0,0,1,1,0,0,0,0,0,0,0,0,0]  & [16-29]     \\  \hline

 Rater2 & if you think it \textcolor{red}{sucks} horribly, that's fine as long as we can fix it 
   &[0,0,0,0,1,0,0,0,0,0,0,0,0,0,0]  &  [16-20]     \\  \hline


 Final Label &  if you think it \textcolor{red}{sucks horribly}, that's fine as long as we can fix it
 
  &[0,0,0,0,1,1,0,0,0,0,0,0,0,0,0]  & [16-29]     \\  \hline

    \end{tabular}

%% file: Sections/evaluation.tex
\section{Evaluation}
\label{sec:evaluation}

\subsection{Evaluation Metrics}
Since our task is based on a sequence tagging approach for toxic spans, we adjusted our evaluation metric from  Martino \emph{et} al.~\cite{da2019fine} that is based on Potthast \emph{et} al.'s plagiarism detection work~\cite{potthast2010evaluation}. Recently, Pavlopoulos \emph{et} al. also used the same metric for toxic span detection in online discussions~\cite{pavlopoulos2022detection}. We decided to use this metric because it provides partial credit for matching the toxic spans inside a sequence. Unlike prior studies~\cite{da2019fine, potthast2010evaluation, pavlopoulos2022detection}, we have chosen token-level comparison instead of character level for measuring the precision, recall, and $F_{1}$ score.  The token-level comparison is taken because token-to-token comparison provides more explainability (\cameraready{comparing the ground truth toxic word to predicted toxic word}) than the character label comparison. For example, a token(s) can represent the toxicity of the whole text, whether a character inside a token does not represent that meaning. 

 Let a code review sample($s$) represent a sequence of tokens $ tok_{0},....{tok_{j}}\subseteq s$ . After IO encoding, the ground truth vector is a sequence of 1's and 0's with 70 values. We calculate the ground truth token offset as $G_{s}= pos_{tok_{m}},....,pos_{tok_{n}}$. So, for each sample($s$), $G_{s}$ contains the position of all toxic tokens ($pos_{tok_{m}}$). When no toxic token exists in the sample, the $G_{s}=empty$.  Similarly, a predictor model predicts the tokens with a floating value. Further, we use a threshold (\cameraready{our experimental evaluation to identify optimal thresholds for each setup is detailed in Section~\ref{sec:threshold-selection}}) to decide whether that token is toxic (1) or non-toxic (0). We generate the predicted token offsets $Pred_{s}$ for each sample from that vector. For better understanding, we put \cameraready{five} examples in table~\ref{tab:prediction} with ground truth (GT) and predicted (Pred) token offsets. Since we processed our text into tokens in preprocessing steps, the first token offset ($tok_{0}$) is for the special token (such as [CLS] for bert tokenizer). Therefore, our first token (i.e., `it') position count starts from 1.



 In the first example of table~\ref{tab:prediction}, we observe that $[7,8,9]$ token offsets are marked as toxic, whereas [7,8,11] offsets are predicted. So, there are two exact matches ($7,8$), one position is not predicted ($9$), and one position is falsely predicted ($10$) as toxic. 
Hence, we used   precision (P), recall (R), and F1 for each sample $s$ are calculated as follows:

\begin{equation}
\label{eqn:precision}
    P^{s}(Pred_{s}, G_{s}) = \frac {|Pred_{s}\cap G_{s}|} {|Pred_{s}|}
\end{equation}

\begin{equation}
\label{eqn:recall}
    R^{s}(Pred_{s}, G_{s}) = \frac {|Pred_{s}\cap G_{s}|} {|G_{s}|}
\end{equation}

\begin{equation}
\label{eqn:f1}
    {F1}^s(Pred_{s}, G_{s}) = \frac {2 * P^{s}(Pred_{s}, G_{s})* R^{s}(Pred_{s}, G_{s})} {P^{s}(Pred_{s}, G_{s}) + R^{s}(Pred_{s}, G_{s})}
\end{equation}

In the equation \ref{eqn:precision}, we define the precision $P^{s}$ for each sample. We define the numerator as the length of the intersection of the set of predicted offsets ($Pred_{s}$) and ground truth token offsets ($G_{s}$). The denominator is the length of predicted offsets ($Pred_{s}$). 
Similarly, we calculate the recall ($R^{s}$) by using equation ~\ref{eqn:recall}. 
 Finally, we combined equation~\ref{eqn:precision} and ~\ref{eqn:recall} to calculate the $F1$ in equation~\ref{eqn:f1}. 

However, these equations can fail due to 0 in denominators. For example, if a model predicts none of the tokens from a sentence belonging to toxic spans, precision is undefined for that sentence. Similarly, for a correctly marked non-toxic instance, recall is undefined. 
We used the same approach as both Pavlopoulos \emph{et} al.~\cite{pavlopoulos2022detection} and the SemEval-2021 Task 5~\cite{pavlopoulos2021semeval} to measure a variation of precision, recall, and \cameraready{F-score} for span detection tasks. In this variation, if the number of predicted toxic tokens is 0 (i.e., $|Pred_{s}| =0$), we check the number of toxic tokens in the ground truth set ($|G_{s}|$). If both sets are empty, the prediction is correct, and we assign this prediction a $precision=1$; otherwise, we assign $precision= 0$. 
On the other hand, if the ground truth set is empty (i.e., $|G_{s}| =0$), we assign $recall =1$ only if the predicted set is also empty (i.e., $|Pred_{s}| =0$), and $recall =0$ otherwise. We would also like to mention that these custom precision/recall measures do not follow traditional precision/recall curve characteristics due to this variation.

We compute and report mean precision, recall, and F-score for the toxic and non-toxic instances separately  since our dataset is highly \cameraready{imbalanced}. In our results, $P_0$, and $P_1$ denote precision for the non-toxic and toxic instances, respectively. 
We consider $F1_{1}$ as our main measure for the experiments because it shows the measurement of the model for the minority (toxic) class tokens.



\begin{table*}
    \caption{{Example of model predictions}. {\textcolor{red}{red}  represents the toxic tokens}}
    \label{tab:prediction}
    \centering

\input{Tables/metric_example}
    \vspace{-14pt}
\end{table*}

To clarify the metric measurement, we show the calculation from the examples of Table~\ref{tab:prediction}. Here, for the first sample, the numerator for equation \ref{eqn:precision} and \ref{eqn:recall} is 2 (i.e., two offsets are intersected). The denominator for equation~\ref{eqn:precision} and \ref{eqn:recall} is 3. So, precision for toxic class ($P_{1}$) is: $\frac{2}{3}= 0.67$, recall for toxic class ($R_{1}$) is: $\frac{2}{3}=0.67$. We calculated the $F_{1}$ as $0.67$. While considering the second sample, the length of ground truth offset $|G_{i}=0|$, but its' predicted offset length $|Pred_{i}=1|$. Since its ground truth is empty, its' metric aligns with the non-toxic class. Hence, equation~\ref{eqn:recall} (recall) would be $\frac{0}{0}$ that would be undefined. For that reason, we put $P_{0}=0$ and $R_{0}=0$ in the second case. Similarly, the third example belongs to the toxic class metric where the length of $|Pred_{i}=0|$. In this case, the equation~\ref{eqn:precision} (precision value) will be $\frac{0}{0}$. For that reason, we set $P_{1}=0$ and $R_{1}=0$ in this case. For the \cameraready{fourth} example, both ground truth and prediction are empty. In those cases, we consider both $P_{0}=1$ and $R_{0}=1$ because we provide full credit for this. \cameraready{The last example shows the measurement where precision and recall are not the same.} 


\subsection{Experimental Setup}
We have done an extensive analysis of each model in our experiment. For accurate estimation of the model performance, we have done 10-fold cross-validation. Using Python's random.seed(), we create stratified 10-folds, which keep a similar ratio between toxic and non-toxic classes for all splits. Further, in each fold, we keep 80\% for the train set, 10\% for the validation set, and the rest 10\% for the test set. We used an NVIDIA Titan RTX GPU with 24 GB memory in Ubuntu 20.04 LTS workstation to conduct the evaluation. 


\subsubsection{Hyperparameters}

We set the following hyperparameters during the training of our model:

\begin{itemize}
    \item \textit{Loss Function:}
We chose a variant of a binary cross-entropy loss function for our task. Since we have multiple tokens (length = 70) with a range of fractional values from 0 to 1, we have added a too-small value (epsilon from Keras) with each prediction. This procedure will help our model to be more stable and prevent the prediction from 0 that can cause  ($\log 0= undefined$) problems. Moreover, we added a clipping between 1 and 0 inside the binary cross-entropy loss to avoid exploding gradients.

\item \textit{Optimizer and Learning Rate:}
We use \textit{Adam} optimizer with a learning rate $1e-5$.


 \item \textit{Number of Epochs:} We set the number of epochs as 30 in each fold. 

 \item\textit{Early Stopping Monitor:}  To prevent the model from overfitting during the training, we set the \textit{EarlyStopping} function from the Keras library~\cite{chollet2015keras} where with the monitor with `val loss’. During training for each epoch, the model is trained with the training dataset and tested with the validation dataset. While the validation loss does not decrease for four consecutive epochs, the model stops training and saves the best model. We also empirically monitored that the optimal `val\_loss' provided the best $F1_{1}$ score for the validation set. 
\end{itemize}

\subsubsection{Threshold Selection} 
\label{sec:threshold-selection}
 
\begin{figure}
	\centering  \includegraphics[width= 9.5cm, height=6 cm]{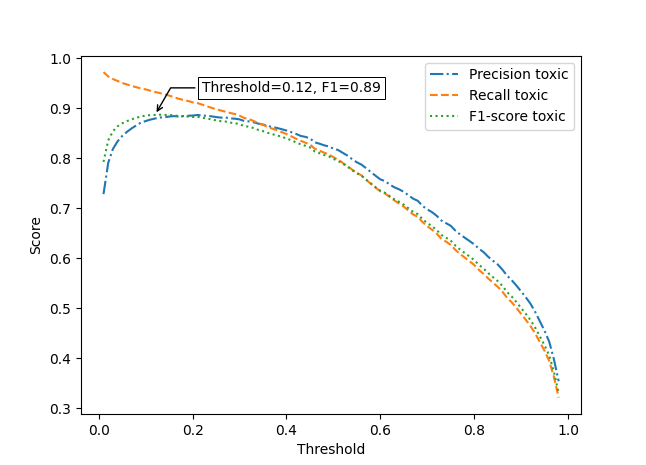}
\caption{Threshold variation for RoBERTa model (\cameraready{Using Validation Set})}
\label{fig:thresh_var}	
	
\end{figure}


Since the models from each fold generate a sequence of 70 length vectors, with a floating value for each token, we need to set a threshold to convert to binary (0 or 1) values. 
In text classification tasks, a threshold of 0.5 is a common choice~\cite{sarker2023automated,pavlopoulos2022detection}.
To identify optimum thresholds, we evaluated our validation set by varying the threshold from 0.01 to 0.99 with a 0.01 increment. We aim to find the threshold resulting in the best $F1_{1}$ score for the validation set. We empirically evaluated each model with this threshold variation for the validation set in 10-fold. With the mean from 10-folds, we found the optimal threshold value for each model to maximize $F1_{1}$ score. For example, the RoBERTa model achieved the best $F1_{1}$ of \cameraready{0.89} with a threshold of \cameraready{0.12 with validation data set}. Figure~\ref{fig:thresh_var} shows performance (precision, recall, and F1) variations for the RoBERTa model against threshold variations \cameraready{using the validation set}. We also noticed that the $F1-score$ for the toxic class remains the same from threshold 0.08 to 0.18 for the RoBERTa model. As we take a \textit{variation} of precision and recall measurement, the plot does not behave like the general characteristics of precision and recall. Figure~\ref{fig:thresh_var} also depicts that by increasing the threshold value, both precision and recall decreased. After calculating the optimal threshold from each model using the validation set, we use that optimal threshold for the corresponding model to predict the test set. During the test set prediction, we also did similar 10-fold cross-validation and got the mean of each metric. Finally, we report the results of each model's performance with the optimal threshold in Table~\ref{tab:results_thresh}.




\subsection{Results with optimal threshold}

\begin{table*}[t]
    \caption{{Experimental Results with the optimal threshold. \cameraready{The runtime of each model and performances during each fold is included in the replication package~\cite{replication}}}}
    \label{tab:results_thresh}
    \centering
    \input{Tables/results_best_threshold}
    \vspace{-12pt}
\end{table*}

We present the results with the optimal threshold for each model in table~\ref{tab:results_thresh}. In the first row, we put the lexicon-based models' performance. Many of the spans in our ground truth contain some specific toxic words. Therefore, the lexicon-based model performed quite well in our study that achieved 0.69 $F1_{1}$ score.
This lexicon-based matching approach also performed better than other transformer models (except the BERT-base model) for non-toxic classes. However, there is a generalizability issue with using the matching approach for detecting toxic spans.

Since our dataset is highly imbalanced,  having a large number of empty spans, 
all of the five transformer models achieved similar scores for $P_{0}$, $R_{0}$, and $F1_{0}$ in the range of $0.90$ $\sim$ $0.95$. For toxic tokens, RoBERTa outperformed the other four models and achieved $0.87$ precision, $0.93$ recall, and $0.88$ $F1_{1}$ score. BERT-base model 
achieved the second best performance with $F_{1}=0.86$. 
DistilBERT and ALBERT models achieved similar performance with $0.85$ $F1_{1}$ score. However, DistilBERT has fewer parameters than other transformer models in our study, and this model is faster during fine-tuning than others. On the other hand, XLNet lacks the performance for both toxic and non-toxic classes than other transformer-based models. 




\begin{boxedtext}
\textbf{Finding 1:} \emph{ While all five transformer-based models achieved better performance than the lexicon-based approach for toxic class, the RoBERTa model outperformed other models with $0.88$ $F1_{1}$} score.
\end{boxedtext}

\subsection{Error Analysis from the best model}

To provide more clarity on our model performance, we have manually analyzed the misclassification with our best-performing model. For that reason, we ran our best-performing RoBERTa model with a threshold of $0.12$ to print misclassification instances. In our final preprocessed dataset, we have a total of 39,438 sentences. During misclassified instance printing, we have done 10 folds. For that reason, we can cover all the samples from our dataset. We have found a total of 3406 ($\sim8.63\%$) sentences where misclassification occurred. However, we categorized the errors into three different types because we are doing a sequence classification problem. Table~\ref{tab:errors} depicts some examples of errors from our model where the first column shows the error types, the second column is for ground truth (GT) token span offsets, and the third column is for predicted token span offsets.


\subsubsection{Partial Disagreement}
 Since we are giving partial credit for the sequence classification metric, we decide to formulate a new error category as Partial Disagreement (PD). We consider an error as PD where the ground truth span offset has some values and the predicted span has some value with some disagreements. We found a total of 945 sentences ($2.4\%$ of the total sample \cameraready{and $27.75\%$ of the error sample}) in this category. The first three examples of Table~\ref{tab:errors} represent the PD category. In the first example, we can see that our rater labeled the `what the hell' phrase inside the toxic span whereas the model predict `the hell' as toxic. 

\subsubsection{False Positives}

We consider False Positives (FP) when a sentence has no spans in its ground truth label but some of its portion is predicted as toxic. In our evaluation, a total of 2226 FPs ($5.64\%$ of total samples, \cameraready{and $65.35\%$ of the error sample}) occurred. We can see three examples of FPs in Table~\ref{tab:errors}. In those examples, the tokens (\textcolor{blue}{FC, stupid, and WTF}) do not represent toxic meanings for those texts. In the third example of FP, the `WTF' represents a library of Linux, not a `what the fuck' phrase. Similarly, the `stupid' word has been used by the reviewer to him/herself. For that reason, that sentence does not have any toxic meaning.

\subsubsection{False Negatives}
We consider the occurrence of False Negatives (FN) where the sentence has single/multiple toxic span offsets but the model predicts no span. The high number of FNs would cause a serious problem for the user of this model because it will miss many toxic instances. Our model has a low amount of FNs where it can not predict toxic span for 235 sentences ($<1\%$ of our total samples, \cameraready{and $6.90\%$ of the error sample}). 
The last two examples on the table~\ref{tab:errors} are FNs that contain some rare toxic phrases (i.e., `Evil', `brain is deficient'). For that reason, our classifier could not predict them as toxic.

\begin{table*}
    \caption{{Example of some errors}}
    \label{tab:errors}
    \centering

\input{Tables/errors}
\end{table*}

\begin{boxedtext}
\textbf{Finding 2:} \emph{ False Positives instances dominated the list of misclassifications. Our models' reliable performances can be attributed to lower instances of `Partial Disagreements' and `False Negatives'.} 
\end{boxedtext}

%% file: Tables/metric_example.tex
  \begin{tabular}{|p{5.6 cm}|p{5.6cm}|p{1.5cm}|p{1.5cm}|p{0.6cm}|p{0.5cm}|} \hline

  \centering{\textbf{Ground Truth Text}}&\centering{\textbf{Predicted Text}} & \textbf{GT Offset} & \textbf{Pred Offset} & \textbf{P} & \textbf{R}\\ \hline

 it is not clear in code \textcolor{red} {what the hell} rest means & it is not clear in code \textcolor{red} {what the} hell rest \textcolor{red}{means}
  &[7,8,9]  & [7,8,11]  & 0.67 & 0.67   \\  \hline

   This will become a trash quick with such a generic name. & This will become a \textcolor{red}{trash} quick with such a generic name.
  &[]  & [5]  & 0 &  0 \\  \hline

  Your indentation is \textcolor{red}{messed up} again & Your indentation is messed up again
  &[4,5]  & []  & 0 & 0 \\  \hline

I do the same as you're suggesting in other code & I do the same as you're suggesting in other code  & [] & [] & 1 & 1 \\  \hline

\cameraready{\textcolor{red}{Oh, shit}, you're right}  & \cameraready{Oh, \textcolor{red}{shit}, you're right}  & [1,2,3] & [3] & 1  & 0.33\\  \hline

    \end{tabular}

%% file: Tables/results_best_threshold.tex

\begin{tabular}{|p{2 cm}|C{2.3 cm}|C{1.7 cm}|C{1.7 cm}|C{1.7 cm}|C{1.7 cm}|C{1.7 cm}|C{1.7 cm}|}
\hline
 
\multirow{2}{*}{\textbf{Models}} & \multirow{2}{*}{\textbf{Optimal Threshold}} &

\multicolumn{3}{c|}{\textbf{Non-toxic words}} & \multicolumn{3}{c|}{\textbf{Toxic words}}  \\ \cline{3-8}

  
  
 
 &&   \textbf{$P_0$} & \textbf{$R_0$} & $F1_0$ & \textbf{$P_1$} & \textbf{$R_1$} & $F1_1$   \\ \hline

Lexicon-based& NA & \textbf{0.95}  & \textbf{0.95}  & \textbf{0.95}  & 0.75  &  0.67  & 0.69       \\ \hline

  BERT-base& 0.15 & \textbf{0.95}  & \textbf{0.95}  & \textbf{0.95}   & \textbf{0.87}  & 0.89  & 0.86        \\ \hline
 RoBERTa& 0.12 &0.92 & 0.92 &0.92   & \textbf{0.87}  & \textbf{0.93}  &  \textbf{0.88}  \\ \hline
 
 DistilBERT&0.17 & 0.94 & 0.94  & 0.94   & 0.85  & 0.89 & 0.85        \\ \hline
 
 ALBERT & 0.11 & 0.92 & 0.92 & 0.92  & 0.85  & 0.89   & 0.85     \\ \hline
  XLNet& 0.10 & 0.90  & 0.90  & 0.90   & 0.79     & 0.88     & 0.81        \\ \hline


\end{tabular}

%% file: Tables/errors.tex
  \begin{tabular}{|p{1cm}|p{1.2cm}|p{1.2 cm}|p{6.2cm}|p{6.2cm}|} \hline

 {\textbf{Error Types}}& \textbf{GT Span} & \textbf{Predicted Span} & {\textbf{Actual Text}} & {\textbf{Predicted Text}} \\ \hline

PD & [14,15,16] & [15,16] &rest seems like too generic name and it's not clear in code \textcolor{red}{what the hell} rest means. & rest seems like too generic name and it's not clear in code what \textcolor{red}{the hell} rest means. \\ \hline

PD& [1] &[1,2]&\textcolor{red}{Damn} grammar :-P & \textcolor{red}{Damn grammar} :-P     \\  \hline

PD &[1,2]& [1,2,3,4,5]& \textcolor{red}{O crap}, hate me: do we still need this one?.
  PD & \textcolor{red}{O crap, hate me}: do we still need this one?.   \\  \hline

  FP & [] & [1] & FC related code should be removed. & \textcolor{red}{FC} related code should be removed.  \\  \hline
  FP & [] & [1] & stupid design on my part. & \textcolor{red}{stupid} design on my part. \\ \hline

FP & [] & [7]& As far as I understood, WTF::HashMap does'nt support it. &As far as I understood, \textcolor{red}{WTF}::HashMap does'nt support it. \\ \hline

 FN & [1] & [] & \textcolor{red}{Evil} spaces must die. & Evil spaces must die. \\ \hline
FN & [2,3,4] &[] & Your \textcolor{red}{brain is deficient}, please fix, also done. & Your brain is deficient, please fix, also done. \\ \hline

    \end{tabular}
    \vspace{-12pt}

%% file: Sections/discussion.tex
\section{Discussion} \label{discussion}
\noindent \textbf{Lesson \#1: Toxic span selection is a highly subjective task for annotators:}
After the initial labeling of the toxic spans, we found that two of our raters showed at least partial disagreement for 928 samples. Human raters do not agree with all samples in selecting the toxic spans. In Some cases, both annotators select the profane words, but one may miss the associated words. For example,\textit{``doesn't this just mean we fucked up the mips syscall.S in bionic?''} text where first labeler marked \textit{``fucked up''} as toxic span and second annotators marked only \textit{``fucked''} as toxic span. In some of the other cases, self-directed anger words such as `argh' or `damn' were mislabeled. Therefore, for similar labeling tasks, we would recommend spending time building a rubric and agreed-upon understandings among the annotators to achieve high IAA.

\noindent \textbf{Lesson \#2: Lexicon-based approach performs well but does not provide generalizability:} We found that our lexicon-based matching approach achieved 0.69 $F1_{1}$ score. Moreover, it performed better for non-toxic classes than the transformer-based supervised training approaches \cameraready{because the lexicon-based approach has less probability of flagging a non-toxic token as toxic (less FPs)}. Though it performed well in our dataset, using this model for a new software engineering dataset may cause serious threats. 
This approach is just token-matching and will miss the associated toxic tokens. Moreover, some tokens do not always represent toxicity. For example,  in \textcolor{red}{\textit{I will kill you}}, where  \textcolor{red}{kill} is toxic. But in  \textcolor{blue}{\textit{Make sure you kill the process first}}, here \textcolor{blue}{kill}   is not toxic. For that reason, the lexicon-based approach may generate a large  number of FPs.



\noindent \textbf{Lesson \#3: Transformer-based models are reliable and explainable for the FOSS community:} In our extensive evaluation, we found that the RoBERTa model outperformed others by achieving $0.88$ $F1_{1}$ score while the other three transformer models also performed well for the toxic class. Since the sequence tagging approach is a challenging task for a new domain, our model can be used by the project maintainers to flag the toxic portion of a text. Moreover, since our best model has fewer false negative cases, FOSS maintainers can use this tool to detect the actual toxic segment from a toxic comment. Apart from that, we have used friendly post-processing, which provides an output with tagging: ``\textcolor{blue}{you're  not  talking  about  neutron,  (\textless toxic\textgreater)  shut up (\textless/ toxic\textgreater)}''.



\noindent \textbf{Lesson \#4: Proactive toxic prevention tool development:}
Since \textit{ToxiSpanSE} is highly precise in identifying toxic excerpts, it is possible to leverage this model to proactively discourage toxic texts. For example, a Gerrit code review plug-in can be developed that highlights toxic excerpts similar to grammatical mistakes or typos, while a review is being written. Such highlights will make an author aware of potential toxic interpretations and may initiate a self-reflection. 

Although the project maintainers would decide on content moderation, they can use our work to develop a tool to rephrase the toxic content to civil comments. Prior work introduced this concept for online communication text~\cite{pavlopoulos2022detection}. The research community from the SE domain and FOSS maintainers may think of this step to reduce the toxic comments from developers' communication.

%% file: Sections/threats.tex
\section{Threats to Validity} \label{threats}
\noindent\emph{A. Internal Validity:} Our selection of code review dataset from a prior work~\cite{sarker2023automated} remains a threat to validity. Biases in the curation of this dataset propagate to our study as well. However, Sarker \emph{et} al.~\cite{sarker2023automated}'s dataset remains the largest labeled toxicity dataset for the SE domain, and it was curated using stratified sampling criteria to span various toxic instances. Since this selected dataset contains only code review comments, it may not adequately represent various other categories of developer communications such as issue discussions or technical question answering. However, that threat may be minimal as we focus on toxic phrases separate from a text's technical contents.

\vspace{2pt}
\noindent \emph{B. Construct Validity:} 
Annotator bias during manual labeling is a potential threat to validity. To mitigate this threat, we reused an already established 
rubric~\cite{sarker2023automated}, 
used a gender-diverse group of annotators including one woman and one man and arranged a discussion with the annotators to build a shared understanding of the rubric before starting 
the annotation process. Moreover, we followed 
recommended practices of independent labeling and 
conflict resolution through discussions. A high value
of Krippendorff's $\alpha$ (i.e., -- 0.81, `almost 
perfect agreement') indicates the reliability of our 
labeling process.

\cameraready{We followed the definition and rubric of toxicity established by Sarker \textit{et} al.~\cite{sarker2023automated}. While Sarker \emph{et} al's conceptualization of toxicity is similar to the ones proposed by Raman \emph{et} al.~\cite{ramanstress} and Miller \emph{et} al.~\cite{miller2022did}, there are subtle differences between their rubrics and ours. Therefore, models trained using our dataset may have degraded performance on datasets released by other studies. 
Similarly, our models may encounter degraded performance on SE datasets of other anti-social communication, such as incivility~\cite{ferreira2021shut} and destructive criticism~\cite{gunawardena2022destructive}. However, this limitation does not apply to our tool pipeline, and it can be retrained to fit other conceptualizations.}

\vspace{2pt}
\noindent\emph{C. External Validity:} Our dataset includes code review comments from four FOSS projects using Gerrit. While we do not have any evidence suggesting the code review interactions on Gerrit are different from other review platforms, such as GitHub pull requests, Phabricator, CodeFlow, and Critique. Our dataset may not adequately represent communication on those platforms. Similarly, as ToxiSpanSE is trained on code review comments, it may have degraded performance on other SE datasets, such as issue discussion, app reviews,  and technical question answering. However, this limit does not apply to our approach, and using a dataset curated from other sources, ToxiSpanSE can be retrained to develop context-specific detectors.

\vspace{2pt}
\noindent\emph{D. Conclusion Validity:}  Using the position-based metric threatens conclusion validity. To mitigate this threat, we adopted our metrics from prior studies with span detection~\cite{pavlopoulos2022detection, da2019fine}. Moreover, since most of our labeled instances are non-toxic, we separately report the performance measures (i.e., $P$, $R$, and $F1$) for both toxic and non-toxic classes.   

%% file: Sections/conclusion.tex
\section{Conclusion and Future Work} \label{conclusion}
In this work, we introduced \textit{ToxiSpanSE}, a SE domain-specific explainable toxicity detector that, in addition to identifying toxic texts, precisely marks the phrases responsible for this prediction. We trained and evaluated \textit{ToxiSpanSE} using 19,651 Code review comments that were manually annotated to mark toxic phrases. We have fine-tuned five different transformers based on encoders that predict the probability of a word being toxic in a given context. We also empirically identified optimum probability thresholds for each of the five models. Our evaluation found a RoBERTa model achieving the best performance with 88\% $F1_{1}$ score. We have made our dataset, scripts, and evaluation results publicly available at \url{https://github.com/WSU-SEAL/ToxiSpanSE}. In addition to facilitating finer-grained toxicity analysis among SE communication, we hope this tool will motivate explainable models for other SE domain-specific NLP classifiers, such as sentiment analysis and opinion mining.